# Power-law decay in first-order relaxation processes

A. Fondado, J. Mira, and J. Rivas
Departamento de Física Aplicada, Universidade de Santiago de Compostela
E-15782 Santiago de Compostela, Spain

Starting from a simple definition of stationary regime in first-order relaxation processes, we obtain that experimental results are to be fitted to a power-law when approaching the stationary limit. On the basis of this result we propose a graphical representation that allows the discrimination between power-law and stretched exponential time decays. Examples of fittings of magnetic, dielectric and simulated relaxation data support the results.

76.20.+q, 76.90.+d, 75.40.Mg

## Introduction

The analysis of the long-decay behavior in real relaxation processes is a subject of interest beyond the academic domain.[1] They are important in science, technology and engineering. From a general point of view, the relaxation phenomena observed in physics, biophysics, chemistry, materials science, polymer science, electronics, etc., present many similar characteristics.[2] Nevertheless, their analysis is sometimes ambiguous due to the noise, the uncertainty in the asymptotic limit and the relatively short interval of time during which relaxation data are recorded. This, together with the possibility of fitting the experimental results to different models with equivalent precision,[3] makes difficult the identification of the processes involved,[4,5] although alternatives are sometimes proposed to solve these problems.[6] Moreover, in usual relaxation processes the coexistence of several diverse mechanisms takes place,[7] and their identification during the fitting stage would be desirable, in order to separate them from the others. Then, to avoid these problems, it would be interesting to obtain models based on general considerations that contained a reduced number of parameters to fit.

In this context, the magnetic properties of assemblies of magnetic interacting particles have been studied by Monte Carlo simulations,[8] and recently Ulrich *et al*.[9] have found that the relaxation rate of the thermoremanent magnetic moment of such assemblies follows an universal power-law. Depending on the value of the exponent, it is found stretched exponential decay for diluted magnetic particles and algebraic decay for concentred ones. These theoretical predictions have been recently confirmed by measurements of relaxation in granular magnetic films.[10,11]

The evolution towards an equilibrium state in relaxation phenomena is expected to approach a certain stationary regime. Generally the word *stationary* indicates a process described by time-independent parameters. Here we will use such basis to try to obtain the functional form of the time-decay of the relaxation process of a system when approaching the stationary limit.

**The model**

Let a relaxing system be described by a field **X**. Its free relaxation is of first-order when the time derivative of first-order of **X** is only a function of the non-time dependence of **X**,

$$\Theta X - \frac{\partial X}{\partial t} = 0 \qquad (1),$$

where $\Theta$ is any spatial-like operator.

A reasonable assumption is that, at long times, the free evolution of the system becomes independent of the initial conditions, and tends to a stationary process. A simple analogy would be a finite relaxing RC network: it tends to a regime where all the capacitors end by discharging with the same time constant. Then, a scalar magnitude $\psi$, representing any kind of average calculated over the state of the system, can be used to describe it. Let us assume that $\psi$ is monotonously decreasing (positive) and that $\lim_{t\to\infty}\psi(t)=0$. In this stationary limit $\psi$ describes the state of the system and therefore its evolution is described by $\psi(t)$ and its time derivative, $\psi'(t)$. We want to abstract this stationary concept, with no reference to any particular process. Then, this first-order process should

1. be described by adimensional magnitudes.

2. be described by $\psi$ and its time derivative, without explicit presence of time.

If we now consider two times, $t_1$ and $t_2$,

$\psi_1 = \psi(t_1), \psi_2 = \psi(t_2),$

$\psi_1' = \psi'(t_1), \psi_2' = \psi'(t_2),$

the magnitudes

$$\frac{\psi_1}{\psi_2}, \frac{\psi_1'}{\psi_2'}$$

fulfil the previous conditions. Any other adimensional magnitude referring to $t_1$ and $t_2$ would be a function of them. We exclude $t_1/t_2$ because it refers explicitly to time and depends on the time origin.

We arrive then to the fact that a first-order process $\psi(t)$ is stationary if there exists a function $f$ fulfilling

$$\frac{\psi_1'}{\psi_2'} = f\left(\frac{\psi_1}{\psi_2}\right) \tag{2}$$

In order to solve Eq. 2 let us consider two times $t_1(x)$ and $t_2(x)$, depending arbitrarily on a parameter $x$. Deriving with respect to $x$:

$$\begin{aligned}\frac{d}{dx}\frac{\psi_1}{\psi_2} &= \frac{1}{\psi_2^2}\left(\psi_2\psi_1'\frac{dt_1}{dx} - \psi_1\psi_2'\frac{dt_2}{dx}\right) \\ \frac{d}{dx}\frac{\psi_1'}{\psi_2'} &= \frac{1}{\psi_2'^2}\left(\psi_2'\psi_1''\frac{dt_1}{dx} - \psi_1'\psi_2''\frac{dt_2}{dx}\right)\end{aligned} \tag{3}$$

Obviously,

$$\frac{d}{dx}\frac{\psi_1}{\psi_2} = 0 \Rightarrow \frac{d}{dx}\frac{\psi_1'}{\psi_2'} = 0$$

that is,

$$\psi_2\psi_1'\frac{dt_1}{dx} = \psi_1\psi_2'\frac{dt_2}{dx} \Rightarrow \psi_2'\psi_1''\frac{dt_1}{dx} = \psi_1'\psi_2''\frac{dt_2}{dx} \tag{4}$$

If we choose functions $t_1(x)$ anf $t_2(x)$ such that the first equality of Eq. 4 is fulfilled, the second will also be fulfilled. In that case, dividing both members of the equations

$$\frac{\psi_1 \psi_1''}{\psi_1'^2} = \frac{\psi_2 \psi_2''}{\psi_2'^2} \tag{5}$$

We can also choose $t_1$ and $t_2$ independently, therefore each of the members of Eq. 5 must be constant $\lambda$:

$$\frac{\psi \psi''}{\psi'^2} = \lambda \tag{6}$$

In order to solve it we rewrite it

$$\frac{\pm \psi''}{\pm \psi'} = \lambda \frac{\psi'}{\psi} \tag{7}$$

and, integrating

$$\ln|\psi'| = \ln A + \lambda \ln|\psi| \tag{8}$$

with $A$ a positive constant. Then

$$\psi' = \pm A \psi^\lambda \tag{9}$$

In a monotonous decreasing process we must use the negative sign, and this leads to the following solution:

$$\begin{aligned} \psi &= \psi_0 e^{-At} &, \quad \lambda = 1 \\ \psi &= [(\lambda-1)A(t+t_0)]^{-\frac{1}{\lambda-1}} &, \quad \lambda > 1 \end{aligned} \tag{10}$$

Introducing appropriate constants $\psi_0$, $\gamma$ and $\tau$, the second solution can be written as

$$\psi = \psi_0 \left(1 + \frac{t}{\gamma \tau}\right)^{-\gamma} \tag{11}$$

with

$$\gamma = \frac{1}{\lambda - 1} \quad , \quad 0 \leq \gamma \leq \infty \tag{12}$$

Note that the Debye process ($\lambda=1$) is the simplest case of Eq. 2 ($\frac{\psi'_1}{\psi'_2} = \frac{\psi_1}{\psi_2}$) and, for finite $t$, is the limit $\gamma \to \infty$ of Eq. 11:

$$\psi_D = \psi_0 e^{-\frac{t}{\tau}} \tag{13}$$

Also, it is worth mentioning that, from Eq. 5

$$\frac{\psi''_1}{\psi''_2} = \frac{\left(\frac{\psi'_1}{\psi'_2}\right)^2}{\frac{\psi_1}{\psi_2}} \tag{14}$$

Given that the inverse of $f$ exists ($\psi$ and $\psi'$ are strictly monotonous decreasing), the second member is function of $\psi'_1/\psi'_2$:

$$\frac{\psi''_1}{\psi''_2} = g\left(\frac{\psi'_1}{\psi'_2}\right) \tag{15}$$

which can be extended by induction to higher order derivatives (i.e., the processes described by the derivatives are also stationary).

## Proposal of graphical representation

An interesting fact is that these considerations can be used to obtain a useful representation of the relaxation process, in which neither $\tau$ nor $t$ appear. Deriving Eq. 10 with respect to $\ln t$, and expressing the result as a function of $\psi$, we arrive to

$$\frac{d\psi}{d\ln t} = -\gamma\psi\left[1-\left(\frac{\psi}{\psi_0}\right)^{\frac{1}{\gamma}}\right] \tag{16}$$

If $d\psi/d\ln t$ vs. $\psi$ is plotted (with $\psi(0) = \psi_0 = 1$ and $\lim_{t\to\infty}\psi(t) = \psi_\infty = 0$) the whole relaxation process can be viewed in an finite window (Fig. 1), like in the Cole-Cole representation in the frequency domain,[12-14] for example.

As a relevant case, it is worth mentioning that the stretched exponential, frequently used in the analysis of relaxation phenomena

$$\psi_K = \psi_0 e^{-\left(\frac{t}{\tau}\right)^\beta} \tag{17}$$

appears as

$$\frac{d\psi_K}{d\ln t} = \beta\psi\ln\frac{\psi}{\psi_0} \tag{18}$$

In this proposed representation, processes $\psi$ described by Eq. 11 with $\gamma\to\infty$ and $\psi_K$ with $\beta\to 1$ reduce to the Debye process. In cases different enough from this, useful informations can be read from the graphs (Fig. 1, with $\psi_0 = 1$ and $\psi_\infty = 0$):

1. Near origin, power-law procesess have finite slope $-\gamma$, and stretched exponential ones have it infinite at origin.

2. Stretched exponential processes have a minimum at $1/e$, whereas power-law processes have minima at points $\psi|_{\text{mín}}$ that depend on $\gamma$,

$$\psi|_{\text{mín}} = \left(\frac{\gamma}{1+\gamma}\right)^{\gamma}, \tag{19}$$

with value

$$\left(\frac{d\psi}{d\ln t}\right)_{\text{mín}} = \left(\frac{\gamma}{1+\gamma}\right)^{\gamma+1}, \tag{20}$$

giving therefore a criterion to identify, from experimental data, stretched exponential or power-law behaviors.

**Results and discussion**

With the aim of checking the validity of our model we have fitted three relaxation phenomena of different nature: i) magnetic (which is one of the most studied scenarios[4,7]), taken from experiments with polycrystalline samples of magnetite ($Fe_3O_4$), ii) dielectric (widely studied by Jonscher[15]), with commercial samples of the polyacrylic acid *Carbopol 907* (Ref. 16) of interest for pharmaceutical applications, and iii) simulated data.

All the examples are first-order processes: The magnetic relaxation of $Fe_3O_4$ in the temperature window $250 < T(K) < 350$ is governed by the diffusion of vacancies[18] that can be described by Eq. 1 (which in fact is a diffusion-like equation). Our dielectric relaxation is a RC discharge, typically described by equations like Eq. 1. Finally, the equations that give rise to the simulated data, fall also in the case of Eq. 1, as we will show later.

i) Magnetic relaxation:

We measured the magnetic permeability, $\mu_r$, of polycrystalline magnetite. $Fe_3O_4$, after a well-defined demagnetization of the sample (magnetic disaccommodation technique[18]). Two processes are clearly distinguishable (Fig. 2, inset). The first one can be attributed to irreversible movements of the domain walls just after demagnetization processes, with topological discontinuities. The second, to reversible displacements after achieving the final topology.[17,19] Both are well fitted with Eqs. 11, with $\gamma$ increasing with temperature in the range 0.2–1 and $\gamma \approx 5$ respectively. Arrhenius fits of $\tau$ (Fig. 2) give similar activation energies (0.9 and 1.0 eV, respectively), that suggest that the damping of the movement of the walls in both processes must be produced by the same mechanism.

ii) Dielectric relaxation

The model was checked with real data taken from the dielectric relaxation at 270 K of Carbopol 907, where $\psi$ is the potential difference between faces after application of an electric current pulse. The sample was prepared from the raw material, supplied as powder. It was compacted to obtain disks with diameter 13 mm and thickness 1 mm. The faces were polished and painted with a conductive coat of graphite, in order to ensure at them constant potentials. In this way we avoid additional relaxation processes due to the charge redistribution at the surfaces.

The final part of the curve (Fig. 3, inset) (stationary regime) is well fitted with Eq. 26. The best proof for the idoneity of the power law in this case is that, upon variation of the interval $(t_1, t_2)$, the parameters of the fit keep reasonably stable in a relatively significative time interval, with $\gamma \cong 4.6$, in contrast with the results from fits to a stretched exponential (Fig. 3).

iii) Simulated data

We made simulations on simple models, in order to obtain the stationary regime in a reasonable time. The model consists of a lineal chain of $N$ elements $x_i$, that relax through non-linear interaction with nearest neighbours. The elements $x_0$ and $x_{n-1}$, interact among them, therefore the system can be interpreted as a periodic chain of period $N$ or a ring. In each iteration (time increase $\Delta t$), $x_i$ increments in a given quantity

$$\Delta x_i = \Delta t \left[ (x_{i+1} - x_i)|x_{i+1} - x_i|^{\lambda-1} + (x_{i-1} - x_i)|x_{i-1} - x_i|^{\lambda-1} \right] , \quad \lambda \geq 1 \quad (21)$$

or

$$\Delta x_i = \Delta t (x_{i-1} + x_{i+1} - 2x_i)|x_{i-1} + x_{i+1} - 2x_i|^{\lambda-1} , \quad \lambda \geq 1 \quad (22)$$

These equations are the unidimensional and discrete version of the following diffusion-like equations:

$$\nabla \cdot \left( |\nabla x|^{\lambda-1} \nabla x \right) - \frac{\partial x}{\partial t} = 0 \quad (23)$$

and

$$|\nabla^2 x|^{\lambda-1} \nabla^2 x - \frac{\partial x}{\partial t} = 0 \quad (24)$$

that fall within the cases described by Eq. 1.

In order to obtain a reasonable variety of cases, the process starts with random $x_i$ values with gaussian distribution (an example is shown in Fig. 4). Besides this, we also proved "ordered" initializations including periodic sequences of periods $N/2, N/3,...$ Before starting a new process, the total average value is substracted to each element (so that $\langle x_i \rangle = 0$), in order to avoid premature rounding off effects in the procedure.

We choose as magnitude $\psi$ the root mean square value of the $x_i$:

$$\psi = \sqrt{\frac{1}{N}\sum_{i=1}^{N} x_i^2} \quad (25)$$

Simulations were made for $\lambda = 1.0, 1.2, 1.5, 1.8, 2.0, 2.5, 3.0$ and $5.0$, varying $\Delta t$ and $N$ (typically $\Delta t = 0.05$ and $N = 64$). The higher $\lambda$, the slower the process. With $\lambda = 5$ we had to increase $\Delta t$ during the process, keeping $\Delta \psi/\psi$ in reasonable values. Anyway, we checked that with smaller $\lambda$ the result is the same as when keeping constant $\Delta t$ during the process.

In all cases the final regime is of the type described by Eq. 11. We verified this checking that the parameters of the fit to Eq. 26 do not change significantly upon changing the point intervals used. Given the random initial conditions, the amplitude $\psi_0$ and the time $\tau$ needed to achieve the stationary regime are also random. Instead, $\gamma$ depends only on $\lambda$, and is coincident with the value given by Eq. 12 that corresponds to a simple relaxor following Eq. 9, with a deviation below 0.1% (with $\lambda = 1$ it is obtained $\gamma > 100$).

The case $\lambda = 1$ corresponds to a linear relaxation. For $\lambda < 1$ and long times $\Delta x_i \to -2x_i$, and the process enters an oscillatory regime that cannot be regarded as of first-order, which led us to discard such cases.

$\gamma$ is invariant under a change of the time origin, and therefore it could be given a physical meaning. In the simulations, $\gamma_0 = 1/(\lambda - 1)$, corresponding to the stationary process, is characteristic of local interactions. The other (pseudostationary) processes, that depend on the time initialization, have smaller $\gamma$s. This trend in $\gamma$ has been observed in all the simulations, and it points to a link of $\gamma$ with the complexity of the process (i.e., the smaller $\gamma$, the more complex the process).

This interpretation is coherent with the results obtained in $Fe_3O_4$, where the irreversible relaxation after demagnetization is specially complicated, and involves slow diffusion processes in each new domain configuration together with magnetic interaction between domains. The velocity of the slow processes increases with temperature, enabling their coordination, and it leads to the increase of $\gamma$, as expected. When the domain walls arrive to the final topology the coordination is maximum, and the system goes to the stationary regime with the highest $\gamma$.

We conclude then, on the basis of quite general conditions, that the relaxation process of first-order follows a power-law time decay on approaching the stationary limit, which is checked in real as well in simulated data. It is proposed also a graphical representation that allows the view of the whole process in a finite window, independently of time.

## Appendix

*Fits*

The fit to Eq. 11 in the general case with unknown $\psi_0$ and $\psi_\infty$ is done in two steps: first calculating $\psi_0$, $\gamma$ and $\psi_\infty$ by means of

$$\frac{d\psi}{d\ln t} = -\gamma(\psi - \psi_\infty)\left[1 - \left(\frac{\psi - \psi_\infty}{\psi_0}\right)^{\frac{1}{\gamma}}\right] \tag{26}$$

and calculating then $\tau$ with the following fit

$$\frac{\psi'}{\psi_0} = -\frac{1}{\tau}\left(\frac{\psi - \psi_\infty}{\psi_0}\right)^{\frac{\gamma+1}{\gamma}} \tag{27}$$

keeping $\psi_0$, $\gamma$ and $\psi_\infty$ constant.

Analogously, the fit to Eq. 17 is done calculating $\psi_0$, $\beta$ and $\psi_\infty$ by means of

$$\frac{d\psi}{d\ln t} = \beta(\psi - \psi_\infty)\ln\frac{(\psi - \psi_\infty)}{\psi_0} \tag{28}$$

and then calculating $\tau$ with the usual fit, with fixed $\psi_0$, $\beta$ and $\psi_\infty$.

*Calculation of the derivatives*

For the fit indicated by Eq. 26 two procedures were followed.

(a) Least square fit: we took an interval of *n* data consecutive in time t, and they were fitted to a polynomial of grade *K*:

$$P(t) = \sum_{k=1}^{K} a_k (t - \langle t \rangle)^k \tag{29}$$

where $\langle t \rangle$ is the mean value of *t* in such interval. The fit is weighed, using as weigh function ρ the square of a Hamming's window. From the coefficients of the fit

$$\begin{aligned} \psi(\langle t \rangle) &= a_0 \\ \psi'(\langle t \rangle) &= a_1 \\ \frac{d\psi}{d\ln t}(\langle t \rangle) &= \langle t \rangle a_1 \end{aligned} \tag{30}$$

With simulated data, without noise, the best results are attained with $K \geq 3$, $n \geq K + 1$. This procedure works well, even with data unequally spaced in time, and it filters noise, as the results are, in a certain sense, the average of the *n* points.

(b) Convolution: The same data interval is convoluted with a function $\eta$, in order to obtain $\langle t \rangle$ and $\langle \psi \rangle$, and with another $\eta'$ to obtain $\langle \psi' \rangle$. $\eta$ and $\eta'$ are obtained from $\rho$ and $\rho'$ by an orthogonalization procedure with respect to the values of $t$ in the points of the interval (that may not be uniformly spaced). It could be said that the procedure is a discretization of the averages,

$$\langle \psi \rangle = \frac{1}{\int_{t_1}^{t_2} \rho \, dt} \int_{t_1}^{t_2} \rho \psi \, dt$$

$$\langle \psi' \rangle = \frac{1}{\int_{t_1}^{t_2} \rho \, dt} \int_{t_1}^{t_2} \rho \psi' \, dt = \frac{1}{\int_{t_1}^{t_2} \rho \, dt} \left\{ [\rho \psi]_{t_1}^{t_2} - \int_{t_1}^{t_2} \rho' \psi \, dt \right\}$$

(31)

with $\rho(t_1) = \rho(t_2) = 0$.

With exact data, the precision of this procedure was somewhat worse than the previous one, but its sensitivity to noise is lower instead, what makes it useful for the processing of experimental data.

**Figure captions**

Figure1: (Color online) Proposal of graphical representation, comparing power-law curves, with $\gamma$ between 0.1 and $\infty$, with stretched exponential ones with $\beta$=0.5 and 1. The case of $\beta$=1 is the same as for $\gamma \rightarrow \infty$.

Figure 2: (Color online) Inset: Fit, at 305 K, of the magnetic relaxation of $Fe_3O_4$ according to our graphical representation. Note how it allows the identification of two relaxation processes. Main frame: Arrhenius plot of the relaxation times of both processes vs. temperature.

Figure 3: (Color online) Inset: representation of the dielectric relaxation of polyacrylic acid according to our proposal. Note the good fit of the left part of the graph (corresponding to the longer times). Main frame: $\tau$'s obtained after fittings to power-law and stretched exponential equations upon variation of the fitted time interval [$(t_1, t_2)$; $t_2$= 1 s]. In the case of power-law fits, the obtained $\tau$'s show a more constant trend, suggesting their validity.

Figure 4: (Color online) Main frame: Example of simulated relaxation (first model, with $\lambda = 2$, random initialization). For the sake of clarity, not all data points are shown. A case is selected from simulations so that the $\psi_0$ of the final lobe is high enough. Inset: Representation of the process according to our proposal. It allows the identification of four different processes. The parameters of the fits are the following: A: $\psi_0 = 0.144$, $\gamma = 0.9998$, $\tau = 6.45 \cdot 10^3$, $\psi_\infty = 0$; B: $\psi_0 = 0.161$, $\gamma = 0.55$, $\tau = 15.2$, $\psi_\infty = 0.123$; C: $\psi_0 = 0.434$, $\gamma = 0.46$, $\tau = 0.36$, $\psi_\infty = 0.163$; D: $\psi_0 = 0.429$, $\gamma = 0.85$, $\tau = 0.172$, $\psi_\infty = 0.273$. Note that for P1 (the stationary process) $\gamma \cong 1$ ($\gamma=1/(\lambda-1)$).

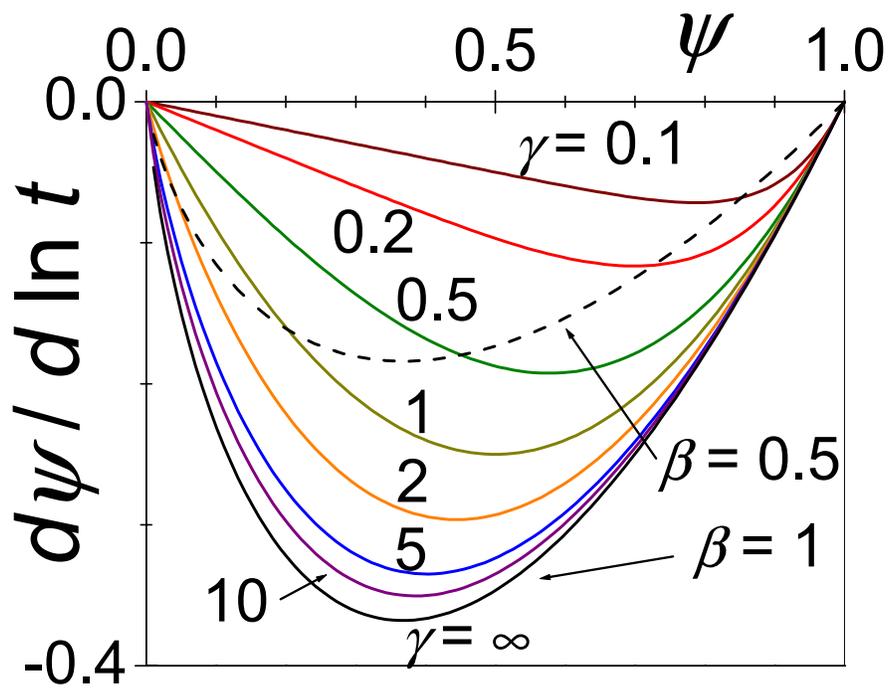

Figure 1

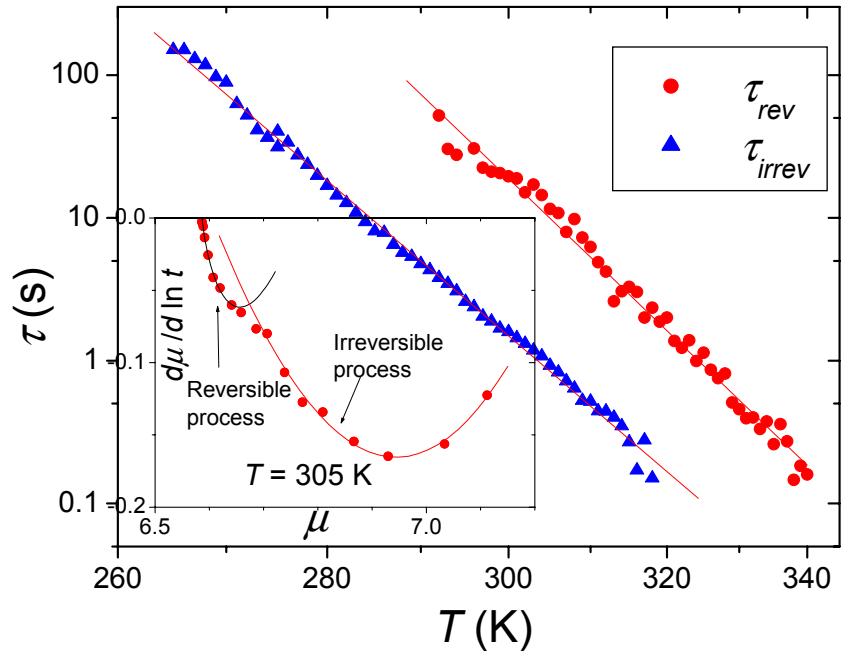

Figure 2

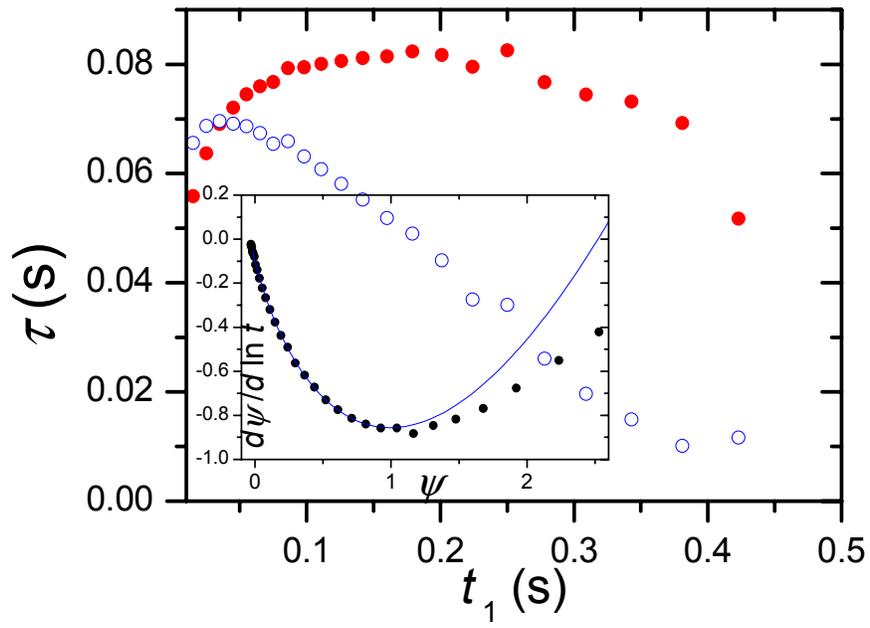

Figure 3

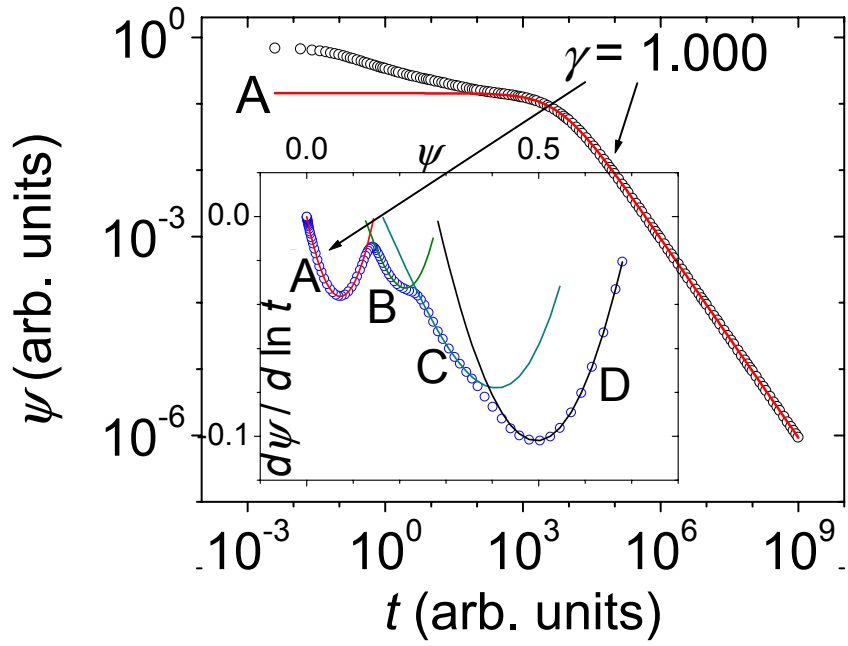

Figure 4